%% file: manuscript.tex
\documentclass{jfm}
\usepackage{epstopdf, epsfig}

\title{End-to-end differentiable learning of turbulence models from indirect observations}

\shorttitle{Differentiable learning of turbulence models from indirect observations}

\shortauthor{C. A. Michelén Ströfer and H. Xiao}

\author{
    Carlos A. Michelén Ströfer\aff{1}\corresp{\email{cmich@vt.edu}}
    \and 
    Heng Xiao\aff{1}\corresp{\email{hengxiao@vt.edu}}
 }

\affiliation{
\aff{1}Kevin T. Crofton Department of Aerospace and Ocean Engineering, Virginia Tech,
Blacksburg, VA 24060, USA
}

\usepackage[colorlinks, citecolor=blue, hypertexnames=False]{hyperref}
\usepackage[dvipsnames]{xcolor}
\usepackage{tikz}
\usetikzlibrary{calc, math, positioning, shapes.arrows, arrows}
\newcommand{\reynoldstress}{\boldsymbol{\tau}} 
\newcommand{\reynoldstresscomp}{\mathsf{\tau}}
\newcommand{\reynoldstressa}{\mathsfbi{a}}

\newcommand{\reynoldstressb}{\mathsfbi{b}}
\newcommand{\reynoldstressbcomp}{\mathsf{b}}
\newcommand{\Tt}{\mathsfbi{T}}
\newcommand{\St}{\mathsfbi{S}}
\newcommand{\Rt}{\mathsfbi{R}}
\newcommand{\It}{\mathsfbi{I}}
\newcommand{\Pt}{\mathsfbi{P}}

\begin{document}

\maketitle

\begin{abstract}
    The emerging push of the differentiable programming paradigm in scientific computing is conducive to training deep learning turbulence models using indirect observations. 
    This paper demonstrates the viability of this approach and presents an end-to-end differentiable framework for training deep neural networks to learn eddy viscosity models from indirect observations derived from the velocity and pressure fields. 
    The framework consists of a Reynolds-averaged Navier--Stokes (RANS) solver and a neural-network-represented turbulence model, each accompanied by its derivative computations. 
    For computing the sensitivities of the indirect observations to the Reynolds stress field, we use the continuous adjoint equations for the RANS equations, while the gradient of the neural network is obtained via its built-in automatic differentiation capability.
    We demonstrate the ability of this approach to learn the true underlying turbulence closure when one exists by training models using synthetic velocity data from linear and nonlinear closures. 
    We also train a linear eddy viscosity model using synthetic velocity measurements from direct numerical simulations of the Navier--Stokes equations for which no true underlying linear closure exists. 
    The trained deep-neural-network turbulence model showed predictive capability on similar flows. 
\end{abstract}


\section{Introduction}
    \label{sec:intro}
    There still is a practical need for improved closure models for the Reynolds-averaged Navier--Stokes (RANS) equations.
    Currently, the most widely used turbulence models are linear eddy viscosity models (LEVM), which presume the Reynolds stress is proportional to the mean strain rate. 
    Although widely used, LEVM do not provide accurate predictions in many flows of practical interest, including the inability to predict secondary flows in non-circular ducts \citep{speziale1982turbulent}. 
    Alternatively, non-linear eddy viscosity models (NLEVM) capture nonlinear relations from both the strain and rotation rate tensors. 
    NLEVM, however, do not result in consistent improvement over LEVM and can suffer from poor convergence \citep{gatski1993explicit}. 
    Data-driven turbulence models are an emerging alternative to traditional single-point closures. 
    Data-driven NLEVM use the integrity basis representation~\citep{pope1975more, ling2016reynolds} to learn the mapping from the velocity gradient field to the normalised Reynolds stress anisotropy field, and retain the transport equations for turbulence quantities from a traditional model. 
    
    It has been natural to train such models using Reynolds stress data \citep[e.g.][]{ling2016reynolds, weatheritt_development_2017}.  
    However, Reynolds stress data from high-fidelity simulations, i.e. from direct numerical simulations (DNS) of the Navier--Stokes equations, is mostly limited to simple geometries and low Reynolds number. 
    It is therefore desirable to train with \emph{indirect observations}, such as quantities based on the velocity or pressure fields, for which experimental data is available for a much wider range of complex flows. 
    Such measurements include full field data, e.g. from particle image velocimetry, sparse point measurements such as from pressure probes, and integral quantities such as lift and drag.  
    Training with indirect observations has the additional advantage of circumventing the need to extract turbulence scales that are consistent with the RANS modelled scales from the high fidelity data \citep[e.g.][]{weatheritt_development_2017}. 
    
    Recently, \citet{zhao2020rans} learned a zonal turbulence model for the wake-mixing regions in turbomachines in symbolic form (e.g., polynomials and logarithms) from indirect observation data by using genetic algorithms. 
    However, while symbolic models are easier to interpret, they may have limited expressive power as compared to, for instance, deep neural networks~\citep{raghu2017expressive}, which are successive composition of nonlinear functions. 
    Symbolic models may therefore not be generalisable and be limited to zonal approaches.  
    More importantly, gradient-free optimisation method such as genetic programming may not be as efficient as gradient-descent methods, and the latter should be used whenever available~\citep{audet2016blackbox}. 
    In particular, deep learning methods~\citep{lecun2015deep} have achieved remarkable success in many fields and represent a promising approach for data-driven NLEVM \citep[e.g.][]{ling2016reynolds}. 
    
    A major obstacle for gradient-based learning from indirect observations is that a RANS solver must be involved in the training and the RANS sensitivites are required to learn the model. 
    While, such sensitivites can be obtained by using adjoint equations, which have long been used in aerodynamic shape optimisation~\citep{jameson1988aerodynamic}, these are not generally rapidly available or straight forward to implement. 
    The emerging interest in differentiable programming is resulting in efficient methods to develop adjoint accompanying physical models, including modern programming languages that come with built-in automatic differentiation~\citep{bezanson2019scientific}, or neural-network-based solutions of partial differential equations \citep[][]{raissi2019physics}. 
    Recently, \citet{holland_field_2019} used the discrete adjoint to  learn a corrective scalar multiplicative field in the production term of the Spalart--Allmaras transport model. 
    This is based on an alternative approach to data-driven turbulence modelling \citep{parish2016paradigm} in which empirical correction terms for the turbulence transport equations are learned while retaining a traditional linear closure (LEVM).  
    
    In this work we demonstrate the viability of training deep neural networks to learn general eddy viscosity closures (NLEVM) using indirect observations. 
    We embed a neural-network-represented turbulence model into a RANS solver using the integrity basis representation, and as a proof of concept we use the continuous adjoint equations to obtain the required RANS sensitivities. 
    This leads to an end-to-end differentiable framework that provides the gradient information needed to learn turbulence models from indirect observations.

\section{Differentiable framework for learning turbulence models}
    \label{sec:training}
\begin{figure}
        \centering
        \input{figure_1.tikz}
        \caption{Schematic of the end-to-end differentiable training framework. 
        The framework consists of two main components, the turbulence model and the observation operator, each of which has a \emph{forward} and \emph{backwards} (adjoint) model. 
        For any value of the trainable parameters $\boldsymbol{w}$ the gradient of the objective function $J$ can be obtained by solving these four problems. 
        The turbulence model consists of a deep neural network representing the closure function using the integrity basis representation $\boldsymbol{\theta}\mapsto\boldsymbol{g}$ and transport equations $\mathcal{T}(k,t_\tau)=0$ for the turbulence quantities. 
        The observation operator consists of solving the RANS equations with the proposed turbulence model and extracting the quantities of interest that are compared to the observations in the cost function $J$.}
        \label{fig:overview}
    \end{figure}
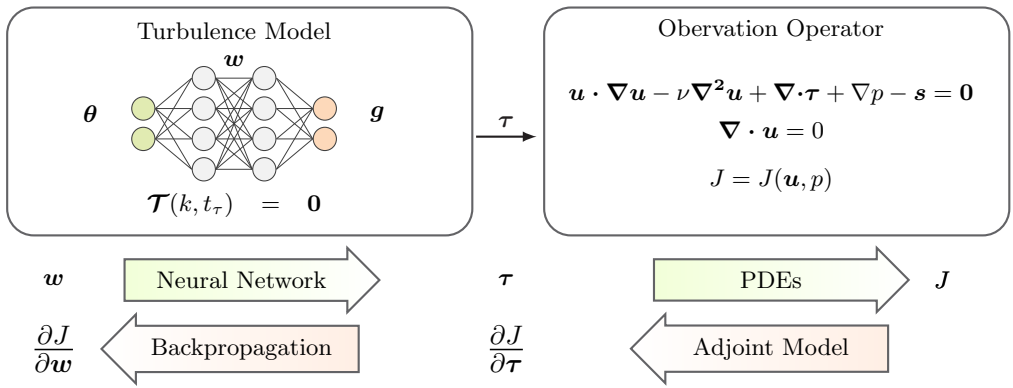

    In this proposed framework a neural network is trained by optimising an objective function that depends on quantities derived from the network's outputs rather than on those outputs directly.
    The training framework is illustrated schematically in figure~\ref{fig:overview} and consists of two components: the turbulence model and the objective function. 
    Each of these two components has a \emph{forward} model that propagates inputs to outputs and a \emph{backwards} model that provides the derivatives of the outputs with respect to the inputs or parameters. 
    The gradient of the objective function $J$ with respect to the network's trainable parameters $\boldsymbol{w}$ is obtained by combining the derivative information from the two components through the chain rule as 
    \begin{equation}
        \frac{\partial J}{\partial \boldsymbol{w}} = \frac{\partial J}{\partial \reynoldstress} \frac{\partial \reynoldstress}{\partial \boldsymbol{w}},
        \label{eq:chain}
    \end{equation}
    where $\reynoldstress$ is the Reynolds stress tensor predicted by the turbulence model. 
      
    \subsection{Forward model}
        For given values of the trainable parameters $\boldsymbol{w}$, the forward model evaluates the cost function $J$, which is the discrepancy between predicted and observed quantities. 
        The forward evaluation consists of two main components: (i) evaluating the neural network turbulence model and (ii) mapping the network's outputs to observation space by first solving the incompressible RANS equations. 

        The turbulence model, shown on the left box in figure~\ref{fig:overview}, maps the velocity gradient field to the Reynolds stress field. 
        The integrity basis representation for a general eddy viscosity model \citep{pope1975more} is given as 
        \begin{equation}
            \reynoldstress = \reynoldstressa + \frac{2k}{3}\It, \qquad \reynoldstressa=2k\reynoldstressb, \qquad \reynoldstressb = \sum_{i=1}^{10}g^{(i)}\Tt^{(i)}, \qquad g^{(i)} = g^{(i)}(\theta_1,\dots,\theta_{5}), 
            \label{eq:reynoldsstress:tau}
        \end{equation}
        where $\reynoldstressa$ is the anisotropic (deviatoric) component of the Reynolds stress, $\reynoldstressb$ is the normalised anisotropic Reynolds stress, $\Tt$ and $\boldsymbol{\theta}$ are the basis tensor functions and scalar invariants of the input tensors, $\boldsymbol{g}$ are the scalar coefficient functions to be learned, and $\It$ is the second rank identity tensor. 
        The input tensors are the symmetric and antisymmetric components of the normalised velocity gradient: $\St=\frac{1}{2}t_\tau\left(\boldsymbol{\nabla u} + \boldsymbol{\nabla u}^\top\right)$ and $\Rt=\frac{1}{2}t_\tau\left(\boldsymbol{\nabla u} - \boldsymbol{\nabla u}^\top\right)$, where $t_\tau$ is the turbulence time-scale and $\boldsymbol{u}$ is the mean velocity. 
        The linear and quadratic terms in the integrity basis representation are given as 
        \begin{equation}
            \begin{array}{c}
            \displaystyle \Tt^{(1)}=\St, \quad
            \Tt^{(2)}=\St\Rt-\Rt\St, \quad
            \displaystyle \Tt^{(3)}=\St^2-\frac{1}{3}\left\{\St^2\right\}\It, \quad
         \displaystyle\Tt^{(4)}=\Rt^2-\frac{1}{3}\left\{\Rt^2\right\}\It \\[8pt]
            \displaystyle \theta_1=\left\{\St^2\right\} 
            \quad \displaystyle\theta_2=\left\{\Rt^2\right\}, 
            \end{array}
            \label{eq:tensorbasis}
        \end{equation}
        where curly braces $\{\boldsymbol{\cdot}\}$ indicate the tensor trace. 

        Different eddy viscosity models differ in the form of the scalar coefficient functions $\boldsymbol{\theta}\mapsto\boldsymbol{g}$ and in the models for the two turbulence scale quantities $k$ and $t_\tau$.
        We represent the scalar coefficient functions using a deep neural network with 10 hidden layers with 10 neurons each and a rectified linear unit (ReLU) activation function.
        The turbulence scaling parameters are modelled using traditional transport equations $\boldsymbol{\mathcal{T}}(k,t_\tau)=0$ with the TKE production term $\Pt$ modified to account for the expanded formulation of Reynolds stress: $\Pt = \boldsymbol{\reynoldstress : \nabla u}$, where $\boldsymbol{:}$ denotes double contraction of tensors. 

        The RANS solver along with its post-processing serves as an observation operator that maps the turbulence model's outputs (Reynolds stress $\reynoldstress$) to the observation quantities (e.g., sparse velocity measurement, drag). 
        This is shown in the right box in figure~\ref{fig:overview}. 
        The first step in this operation is to solve the RANS equations with the given Reynolds stress closure to obtain mean velocity $\boldsymbol{u}$ and pressure $p$ fields.  This is followed by post-processing to obtain the observation quantities (e.g., sampling velocities at certain locations or integrating surface pressure to obtain drag).
        When solving the RANS equations, explicit treatment of the divergence of Reynolds stress can make the RANS equations ill-conditioned \citep{wu2019reynolds,brener2021conditioning}. 
        We treat part of the linear term implicitly by use of an effective viscosity $\nu_\textrm{\tiny eff}$ which is easily obtained since with the integrity basis representation the linear term is learned independently.
        The incompressible RANS equations are then given as
        \begin{equation}
            \begin{array}{c}
                \displaystyle \boldsymbol{u\cdot\nabla u} - \boldsymbol{\nabla\cdot}  \nu_\textrm{\tiny eff}\boldsymbol{\nabla u} - \boldsymbol{\nabla u\cdot} \nabla\nu_\textrm{\tiny eff} + \boldsymbol{\nabla\cdot} \reynoldstressa_\textrm{\tiny NL} + \nabla p^* = \boldsymbol{s},  \\[5pt] 
                \displaystyle \boldsymbol{\nabla\cdot u} = 0,  \\[3pt]
                \displaystyle \nu_\textrm{\tiny eff} = \nu-g^{(1)}kt_\tau, \qquad \reynoldstressa_\textrm{\tiny NL} = 2k\sum_{i=2}^{10}g^{(i)}\Tt^{(i)}, \qquad p^* = p + \frac{2k}{3}
            \label{eq:ransimp}
            \end{array}
        \end{equation}
        where the term $\boldsymbol{\nabla\cdot}\nu_\textrm{\tiny eff}\boldsymbol{\nabla u}$ is treated implicitly.
        Here $\reynoldstressa_\textrm{\tiny NL}$ represents the non-linear component of the Reynolds stress anisotropy, the isotropic component of Reynolds stress is incorporated into the pressure term $p^*$, and $\boldsymbol{s}$ is the momentum source term representing any external body forces. 
        The coefficients $g^{(i)}$ are outputs of the neural network-based turbulence model that have the fields $\boldsymbol{\theta}$ as input.

    \subsection{Adjoint model}
        For a proposed value of the trainable parameters $\boldsymbol{w}$ the backwards model (represented by left-pointing arrows in figure~\ref{fig:overview}) provides the gradient $\partial J/\partial \boldsymbol{w}$ of the cost function with respect to the trainable parameters. 
        This is done by separately obtaining the two terms on the right hand side of equation~(\ref{eq:chain}): (i) the gradient of the Reynolds stress $\partial \reynoldstress/\partial\boldsymbol{w}$ from the neural network turbulence model and (ii) the sensitivity $\partial J/\partial\reynoldstress$ of the cost function to the Reynolds stress, using their respective adjoint models. 
        Combining the two adjoint models results in an end-to-end differentiable framework, whereby the gradient of observation quantities (e.g. sparse velocity) with respect to the neural network's parameters can be obtained.

        The gradient of the Reynolds stress with respect to the neural network's parameters $\boldsymbol{w}$ is obtained in two parts using the chain rule.
        The gradient of the neural network's outputs with respect to its parameters, $\partial g/\partial \boldsymbol{w}$, is efficiently obtained by backpropagation, which is a reverse accumulation automatic differentiation algorithm for deep neural networks that applies the chain rule on a per-layer basis.  
        The sensitivities of the Reynolds stress to the coefficient functions are obtained as $\partial \reynoldstress/\partial g^{(i)}= 2k\Tt^{(i)}$ from differentiation of equation~\ref{eq:reynoldsstress:tau}, which is a linear tensor equation.
            
        For the sensitivity of the objective function with respect to the Reynolds stress we derived the appropriate continuous adjoint equations.
        Since the Reynolds stress must satisfy the RANS equations, this is a constrained optimisation problem. 
        The problem is reformulated as the optimisation of an unconstrained Lagrangian function with the Lagrange multipliers described by the adjoint equations.  
        The resulting adjoint equations are 
        \begin{equation}
            \begin{array}{c}
                \displaystyle \boldsymbol{u\cdot\nabla\hat{u}} + \boldsymbol{\nabla\hat{u}\cdot u} + \nu\boldsymbol{\nabla^2\hat{u}} - \nabla\hat{p}  = \frac{\partial J_\Omega}{\partial \boldsymbol{u}} \\[5pt]
                \displaystyle \boldsymbol{\nabla\cdot \hat{u}} = - \frac{\partial J_\Omega}{\partial p},
            \end{array}
            \label{eq:adjoint}
        \end{equation}
        where $\boldsymbol{\hat{u}}$ and $\hat{p}$ are the Lagrange multipliers, referred to as adjoint velocity and adjoint pressure, respectively, and $J_\Omega$ is the scalar field such that the objective function is given as the integral $J=\int J_\Omega \mathrm{d}\Omega$ over the solution domain $\Omega$. 
        When the cost function is instead given as an integral over the domain boundary (e.g. drag) the cost function affects the boundary conditions of the adjoint equations instead \citep{othmer2008continuous}. 
        When solving the adjoint equation using a segregated approach such as the SIMPLE algorithm used here, the adjoint transpose convection term $\boldsymbol{\nabla\hat{u}\cdot u}$ is treated explicitly and can result in instabilities \citep{oriani2016alternative}. 
        For this reason it is common to dampen or eliminate this term \citep{othmer2014adjoint}, and here we eliminate it. 
        After solving the adjoint equations, the sensitivity of the function $J$ to the Reynolds stress is given as the gradient of the adjoint velocity
        \begin{equation}
            \frac{\partial J}{\partial\reynoldstress}=-\boldsymbol{\nabla\hat{u}}.
            \label{eq:gradrans}
        \end{equation}
        The details of the derivation and use of the continuous adjoint equations are shown in~\citep{michelen2021machine} for interested readers.

    \subsection{Gradient descent procedure}
        \label{sec:training:train}
        The training is done using the Adam algorithm, a gradient descent algorithm with momentum and adaptive gradients commonly used in training deep neural networks. 
        The default values for the Adam algorithm are used, including a learning rate of $0.001$.
        The training requires solving the RANS equations at each training step. 
        In a given training step the inputs $\theta_i$ are updated based on the previous RANS solution and scaled to the range $0$--$1$, and the RANS equations are then solved with fixed values for the coefficient functions $\boldsymbol{g}$. 
        The inputs $\theta_i$ and their scaling parameters are fixed at a given training step and converge alongside the main optimisation of the trainable parameters $\boldsymbol{w}$.
        
        initialisation of the neural network's parameters requires special consideration. 
        The usual practice of random initialisation of the weights is not suitable in this case since it leads to divergence of the RANS solution. 
        We use existing closures (e.g. a laminar model with $g^{(i)}=0$ or a linear model with $g^{(1)}=-0.09$) to generate data for pre-training the neural network and thus provide a suitable initialisation. 
        This has the additional benefit of embedding existing insight into the training by choosing an informed initial point in the parameter space. 
        When pre-training to constant values (e.g. $g^{(1)}=-0.09$) we add noise to the pre-training data, since starting from very accurate constant values can make the network difficult to train.

\section{Test cases}
    \label{sec:cases}
    The viability of the proposed framework is demonstrated by testing on three test cases. 
    The first two cases use synthetic velocity data obtained from a linear and a non-linear closure, respectively, to train the neural network. 
    The use of synthetic data allows us to evaluate the ability of the training framework to learn the true underlying turbulence closure when one exists. 
    In the final test case realistic velocity measurements, obtained from a DNS solution and for which no known true underlying closure exists, are used to learn a linear eddy viscosity model. 
    The trained LEVM is then used to predict similar flows and the predictions are compared to those from a traditional LEVM.

    \subsection{Learning a synthetic LEVM from channel flow}
        \label{sec:cases:linear}
\begin{figure}
            \centering
            \includegraphics[trim=29 0 29 0, clip]{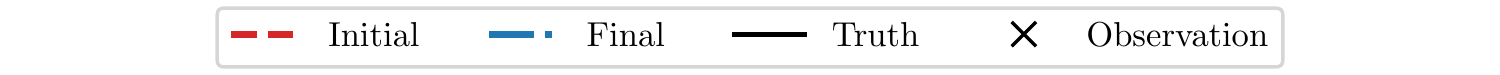} \\
            (\textit{a})\raisebox{-1\height}{\includegraphics[trim=5 0 5 0, clip]{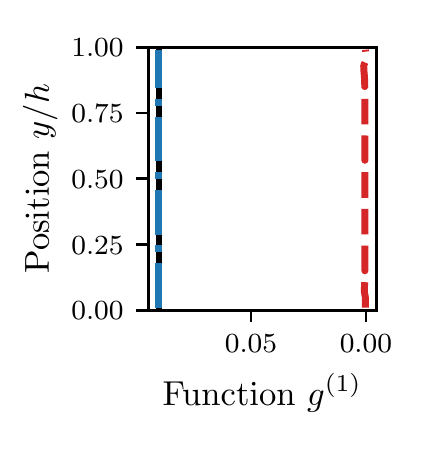}} 
            (\textit{b})\raisebox{-1\height}{\includegraphics[trim=5 0 5 0, clip]{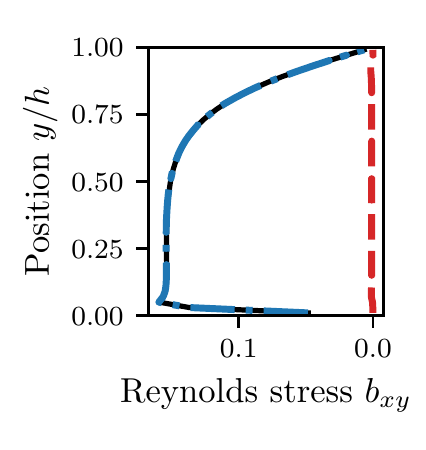}}
            (\textit{c})\raisebox{-1\height}{\includegraphics[trim=5 0 5 0, clip]{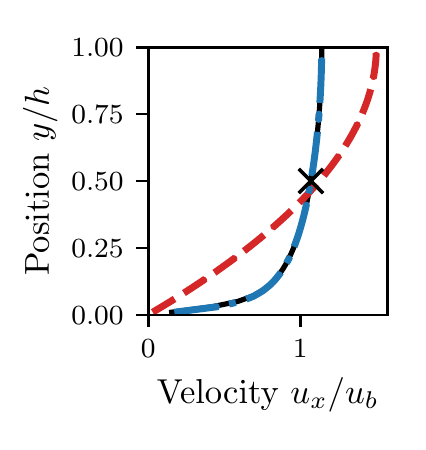}}
            \caption{Results of learning a LEVM from a single velocity measurement in a turbulent channel flow: (\textit{a}) the learned coefficient function, (\textit{b}) the anisotropic Reynolds stress field, and (\textit{c}) the velocity. The final (trained) results overlap with the truth and the two are visually indistinguishable.} 
            \label{fig:channel}
        \end{figure}

        As a  first test case we use a synthetic velocity measurement at a single point from the turbulent channel flow to learn the underlying linear model. 
        The flow has a Reynolds number of $10,000$ based on bulk velocity $u_b$ and half channel height $h$. 
        The turbulent equations used are the $k$--$\omega$ model of \citet{wilcox1998turbulence}, and the synthetic model corresponds to a constant $g^{(1)}=0.09$. 
        For the channel flow there is only one independent scalar invariant and $\Tt^{(1)}$ is the only linear tensor function in the basis.
        We therefore use a neural network with one input and one output which maps $\theta_1\mapsto g^{(1)}$. 
        The network has $1021$ trainable parameters and is pre-trained to the laminar model $g^{(1)}=0$. 
        The sensitivity of the predicted point velocity to the Reynolds stress is obtained by solving the adjoint equations with $J_\Omega$ equal to the velocity field times a radial basis function. 
        Figure~\ref{fig:channel} shows the results of the training. 
        The trained model not only results in the correct velocity field, but the correct underlying model is learned.
        
    \subsection{Learning a synthetic NLEVM from flow through a square duct}
        \label{sec:cases:quadratic}
        As a second test case we use a synthetic full field velocity measurement from flow in a square duct to learn the underlying nonlinear model.  
        The flow has a Reynolds number of $3,500$ based on bulk axial velocity $u_b$ and half duct side length $h$.  
        This flow contains a secondary in-plane flow that is not captured by LEVM \citep{speziale1982turbulent}. 
        For the objective function, $J_\Omega$ is the difference between the measured and predicted fields, with the discrepancy of the in-plane velocity scaled by a factor of $1,000$ as to have a similar weight to the axial velocity discrepancy.  
        The NLEVM is the Shih quadratic model \citep{shih1993realizable} which, using the basis in equation~\ref{eq:tensorbasis}, can be written as
        \begin{equation}
            \begin{array}{ll}
                \displaystyle g_1(\theta_1, \theta_2) = \frac{-2/3}{1.25 + \sqrt{2\theta_1} + 0.9 \sqrt{-2\theta_2}}, & \displaystyle
                g_2(\theta_1, \theta_2) = \frac{7.5}{1000 + (\sqrt{2\theta_1})^3}, \\[10pt]
                \displaystyle g_3(\theta_1, \theta_2) = \frac{1.5}{1000 + (\sqrt{2\theta_1})^3}, & \displaystyle
                g_4(\theta_1, \theta_2) = \frac{-9.5}{1000 + (\sqrt{2\theta_1})^3}. 
            \end{array}
            \label{eq:shih_final}
        \end{equation}
    
        For the flow in a square duct only four combinations of the Reynolds stress components affect the predicted velocity \citep{speziale1982turbulent}: $\reynoldstresscomp_{xy}$ and $\reynoldstresscomp_{xz}$ in the axial equation and $\reynoldstresscomp_{yz}$ and $(\reynoldstresscomp_{zz}-\reynoldstresscomp_{yy})$ in the in-plane equation. 
        In this flow the in-plane velocity gradients are orders of magnitude smaller than the gradients of the axial velocity $u_x$. 
        For these reasons only two combinations of coefficient functions can be learned, $g^{(1)}$ and the combination $g^{(2)}-0.5g^{(3)}+0.5g^{(4)}$, and there is only one independent scalar invariant with $\theta_1\approx-\theta_2$. 
        
        \begin{figure}
            \centering
            (\textit{a}) \raisebox{-1\height}{\includegraphics[trim=10 10 10 0, clip]{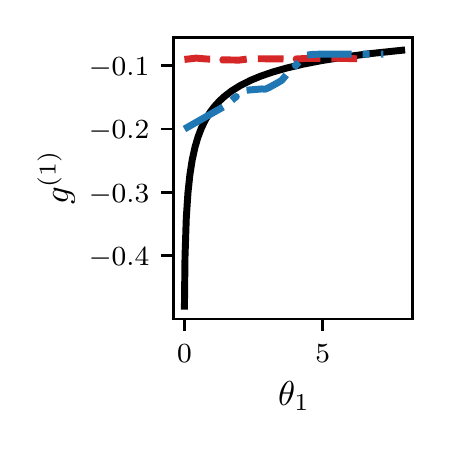}} 
            (\textit{b}) \raisebox{-1\height}{\includegraphics[trim=10 10 10 0, clip]{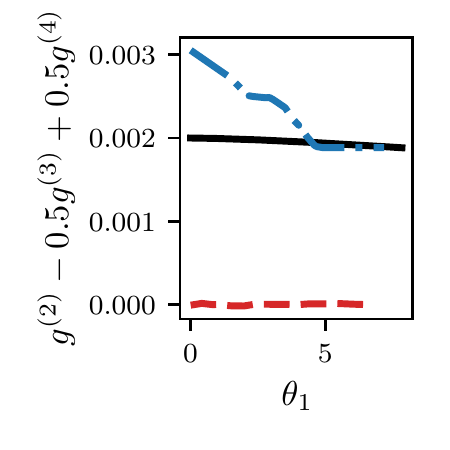}} 
            \raisebox{-1.5\height}{\includegraphics[trim=170 120 170 120, clip]{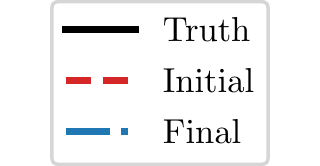}}
            \caption{Results of learning a NLEVM, the Shih quadratic model, from full field velocity measurements in flow through a square duct. The results shown are the two combinations of coefficient functions that have an effect on velocity plotted against the scalar invariant $\theta_1\approx-\theta_2$.}
            \label{fig:duct_g}
        \end{figure}
        
        \begin{figure}
            \centering
            (\textit{a})\raisebox{-1\height}{
            \begin{tabular}{cccccc} 
                & $u_x/u_b$ & $u_y/u_b$ & $\reynoldstressbcomp_{xy}$ & $\reynoldstressbcomp_{yz}$ & $(\reynoldstressbcomp_{yy}-\reynoldstressbcomp_{zz})$
                \\
                & & $\times10^{3}$ & $\times10^{1}$ & $\times10^{3}$ & $\times10^{2}$ 
                \\
                \rotatebox[origin=c]{90}{Initial} & 
                \raisebox{-.5\height}{\includegraphics[scale=0.35, trim=10 0 10 0, clip]{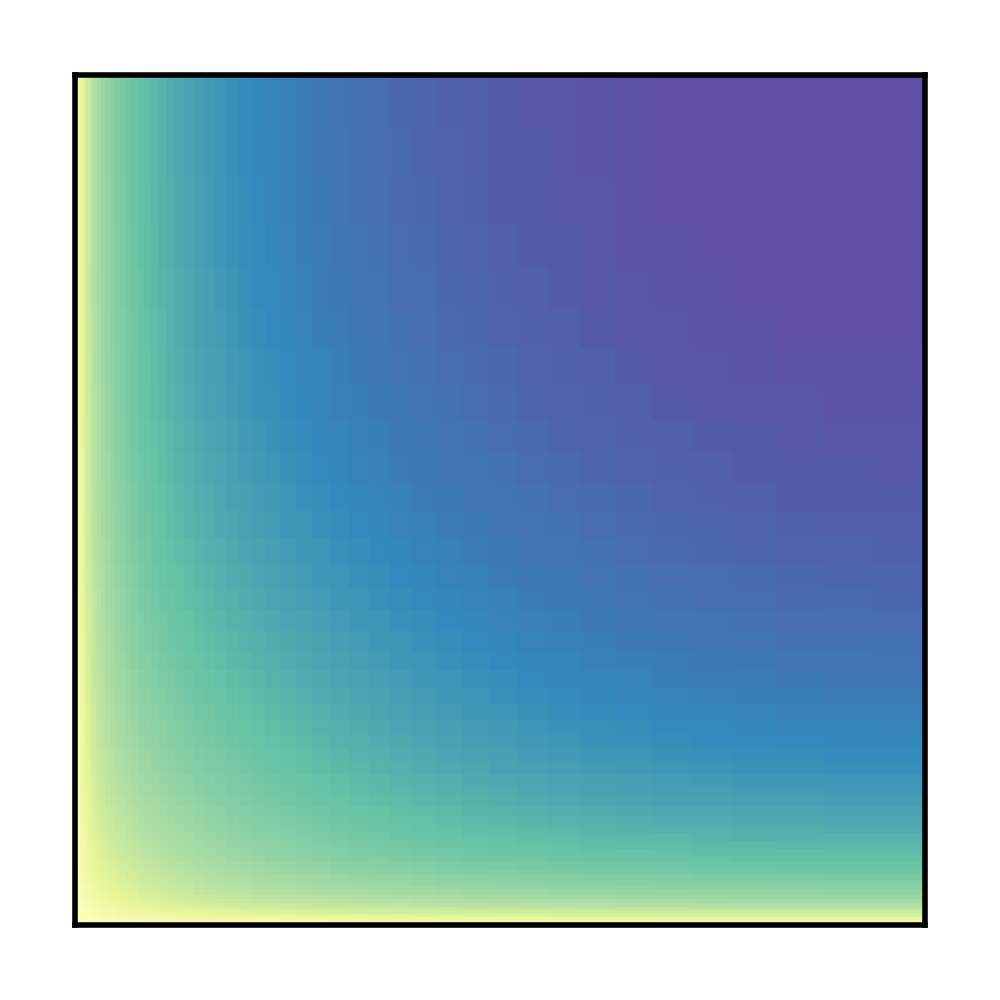}} & 
                \raisebox{-.5\height}{\includegraphics[scale=0.35, trim=10 0 10 0, clip]{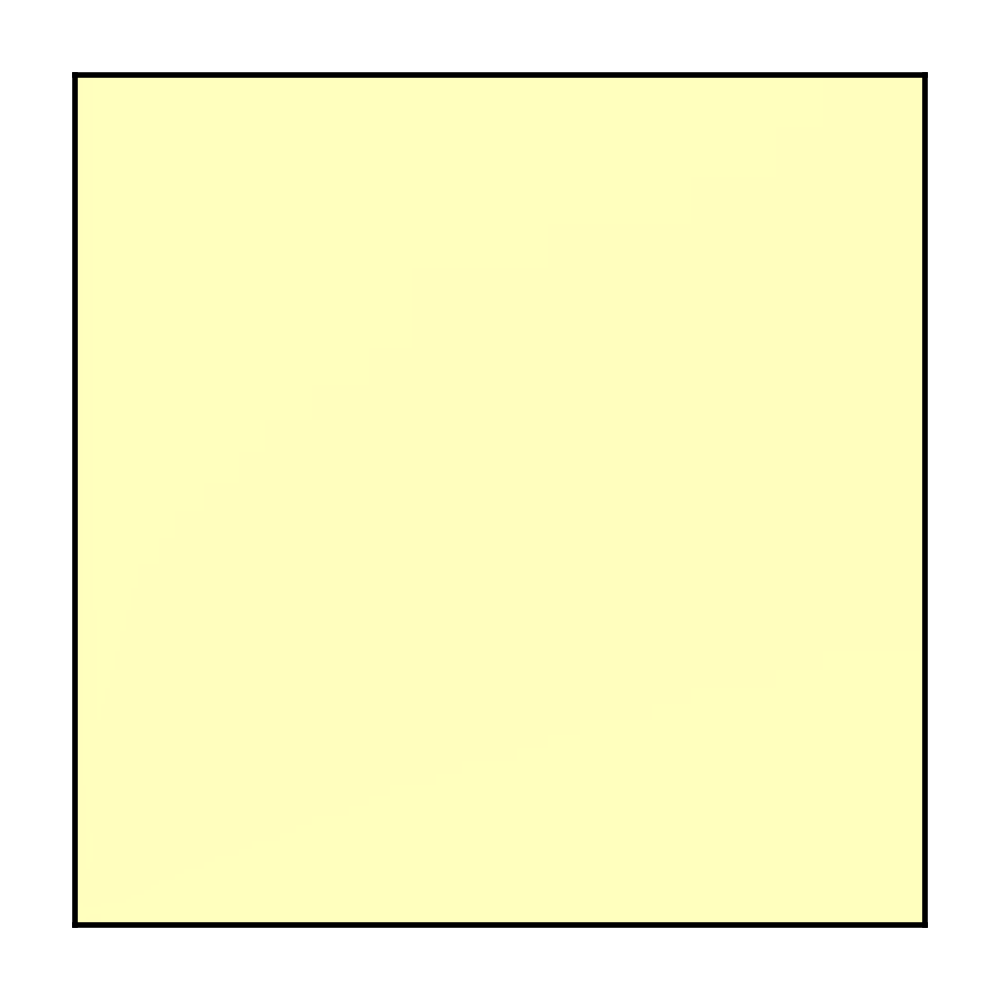}} & 
                \raisebox{-.5\height}{\includegraphics[scale=0.35, trim=10 0 10 0, clip]{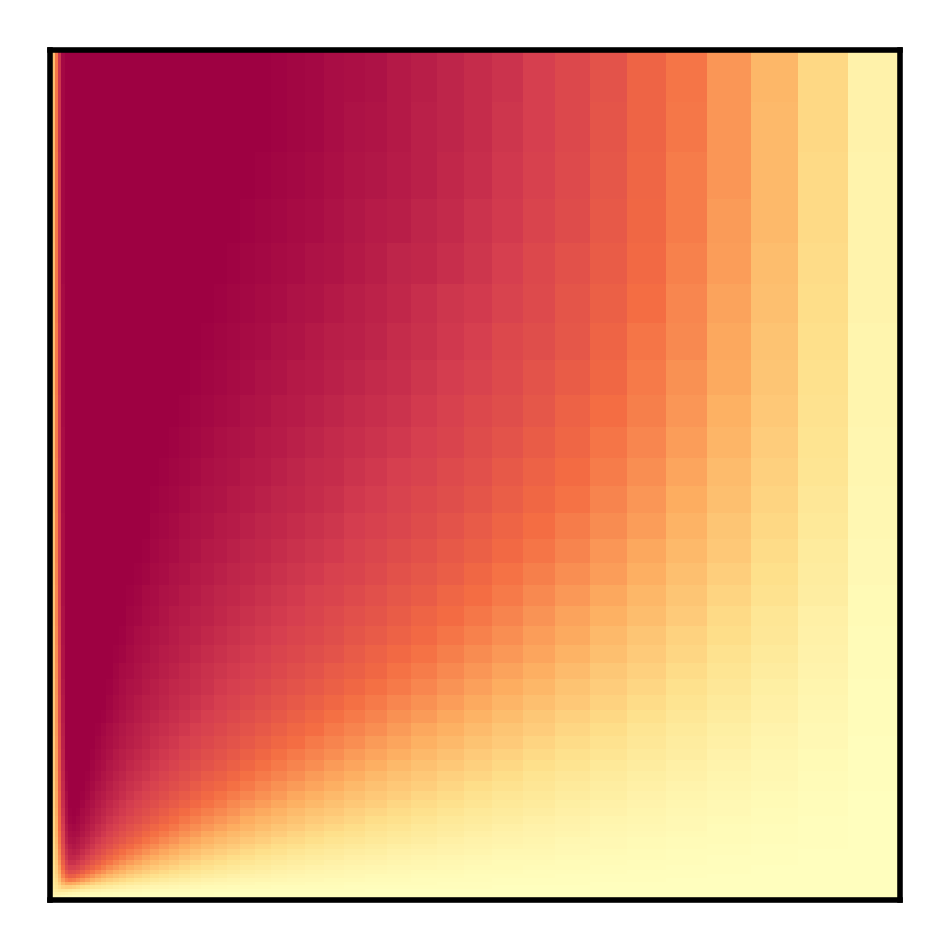}} & 
                \raisebox{-.5\height}{\includegraphics[scale=0.35, trim=10 0 10 0, clip]{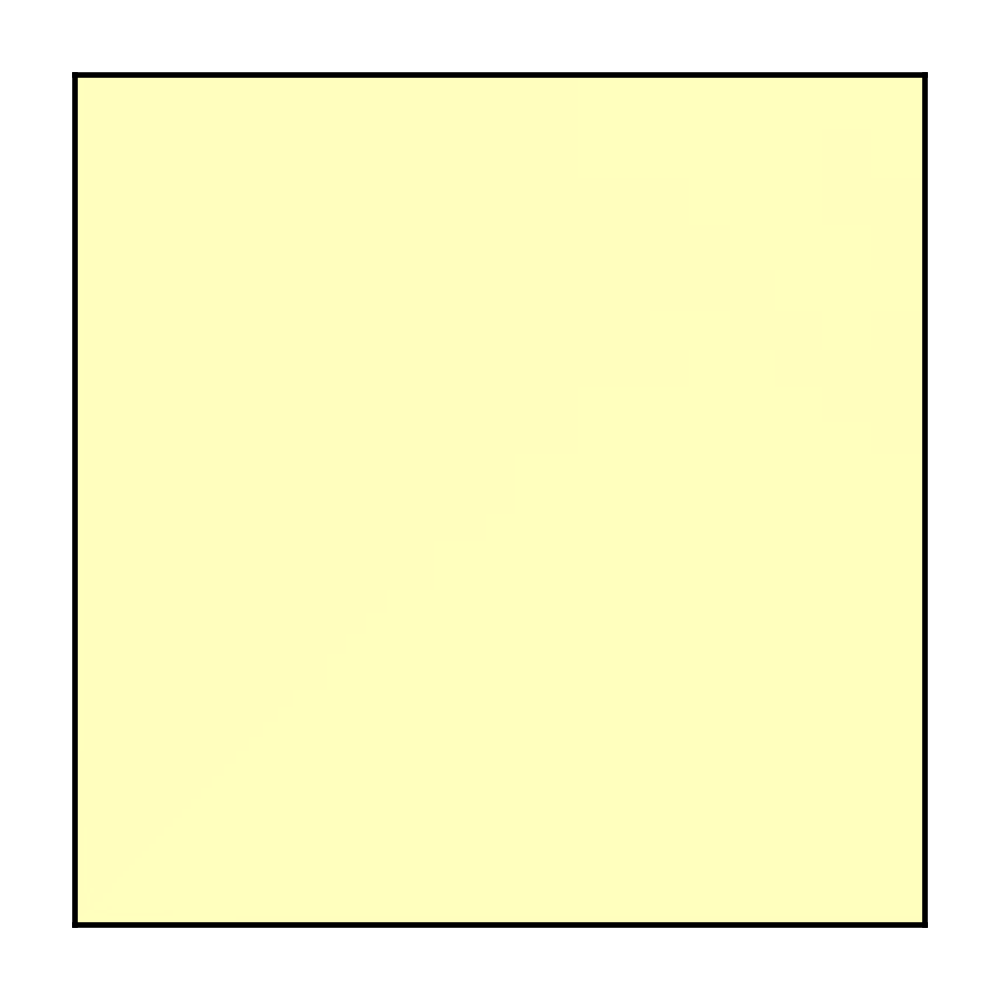}} & 
                \raisebox{-.5\height}{\includegraphics[scale=0.35, trim=10 0 10 0, clip]{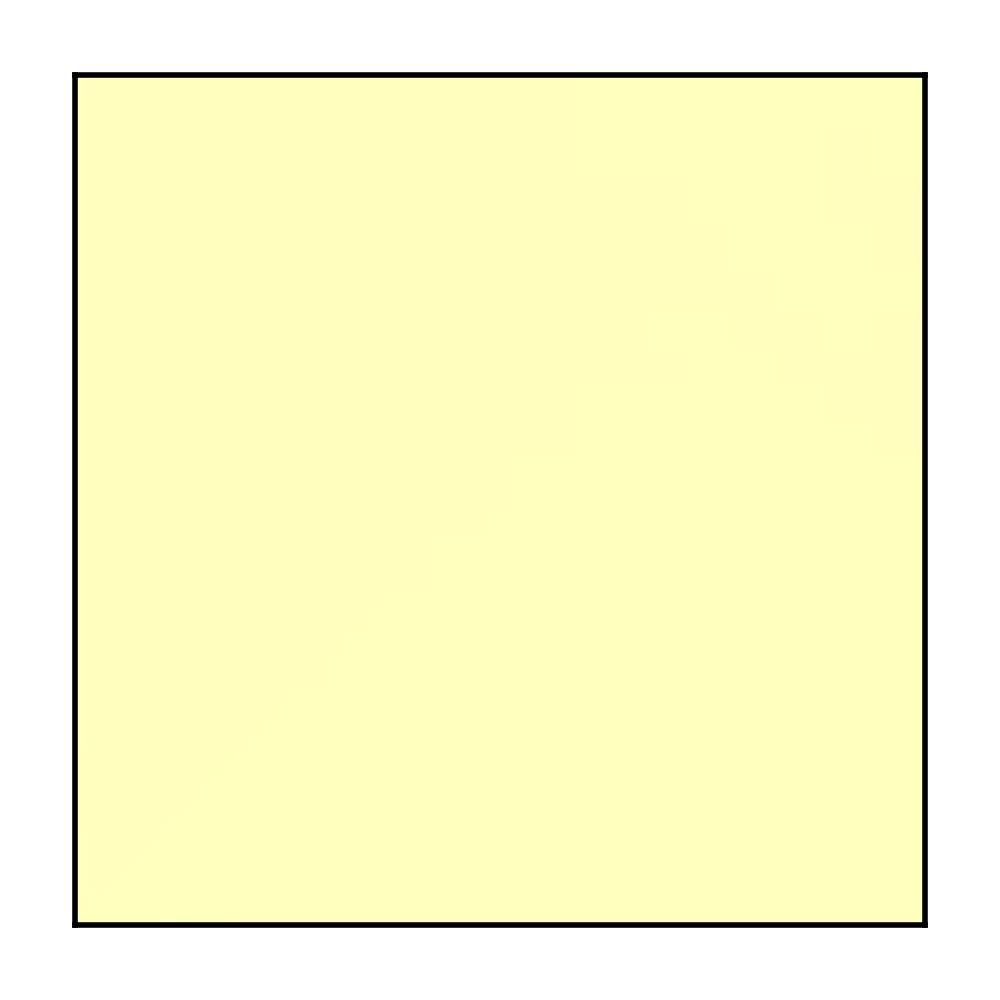}} 
                \\
                \rotatebox[origin=c]{90}{Final} & 
                \raisebox{-.5\height}{\includegraphics[scale=0.35, trim=10 0 10 0, clip]{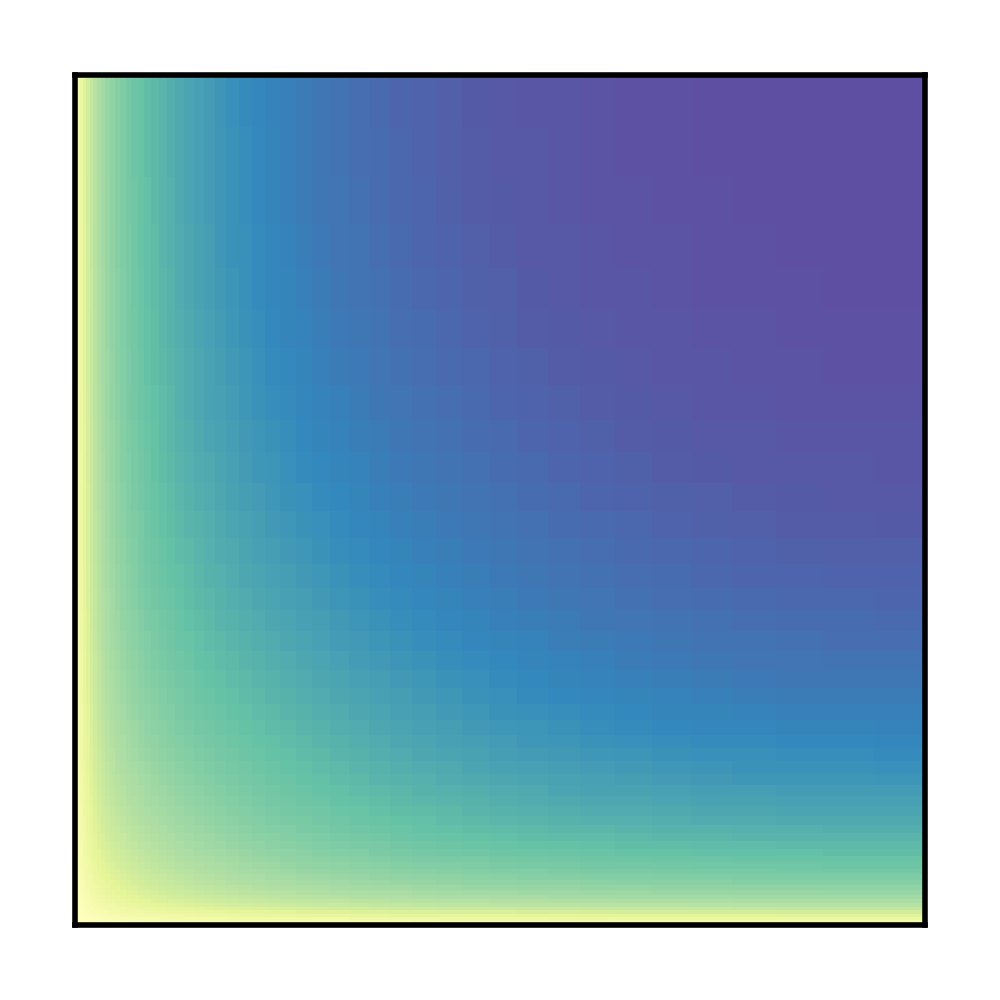}} & 
                \raisebox{-.5\height}{\includegraphics[scale=0.35, trim=10 0 10 0, clip]{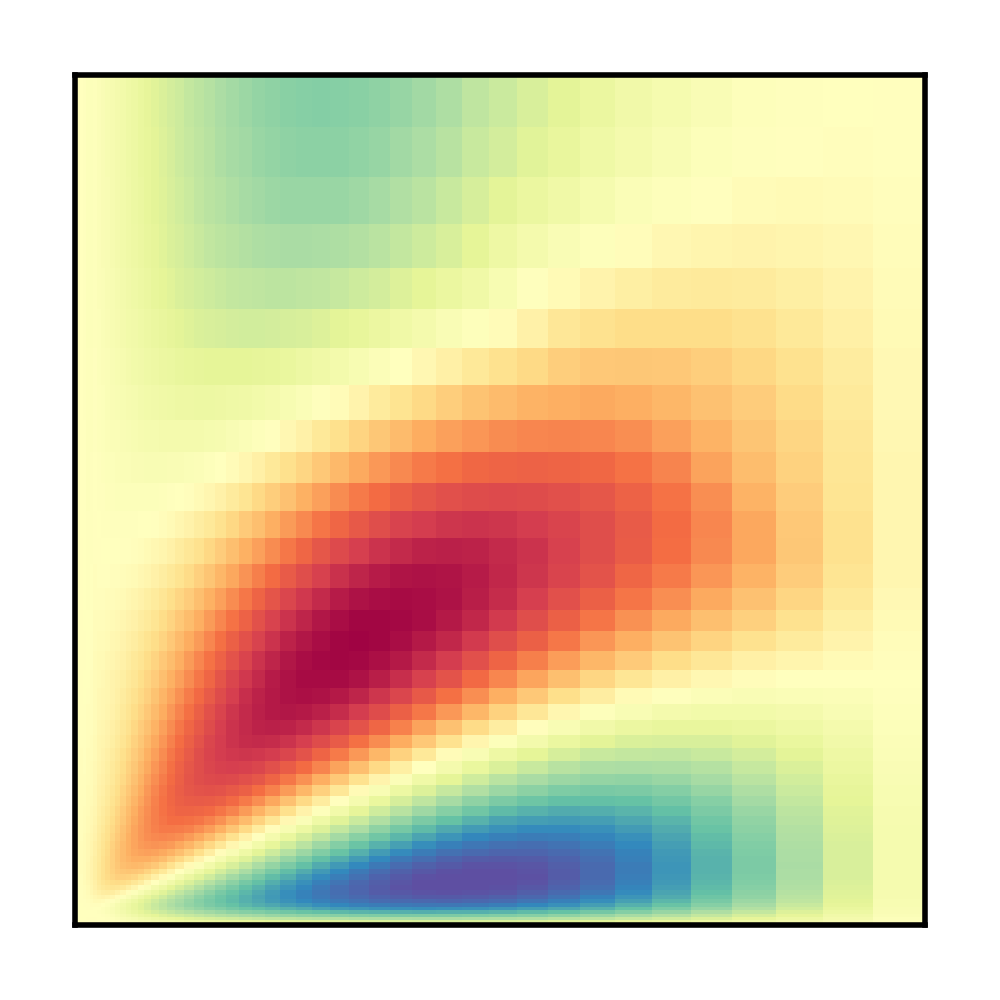}} & 
                \raisebox{-.5\height}{\includegraphics[scale=0.35, trim=10 0 10 0, clip]{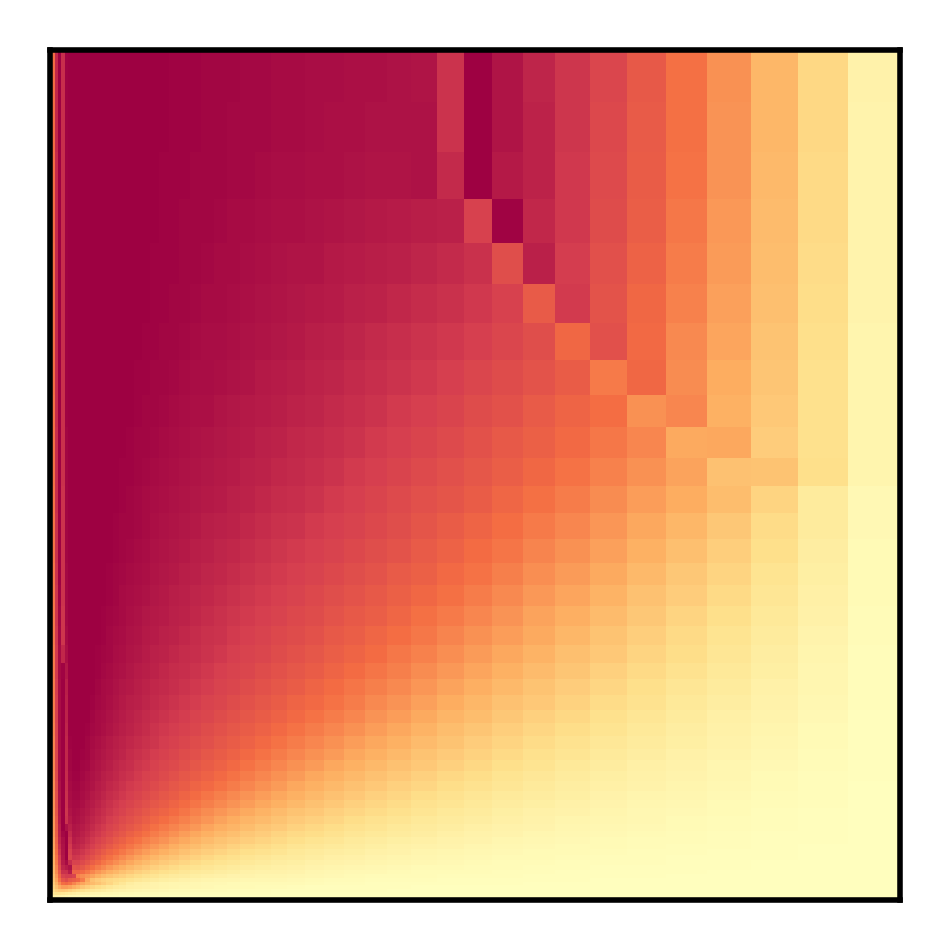}} & 
                \raisebox{-.5\height}{\includegraphics[scale=0.35, trim=10 0 10 0, clip]{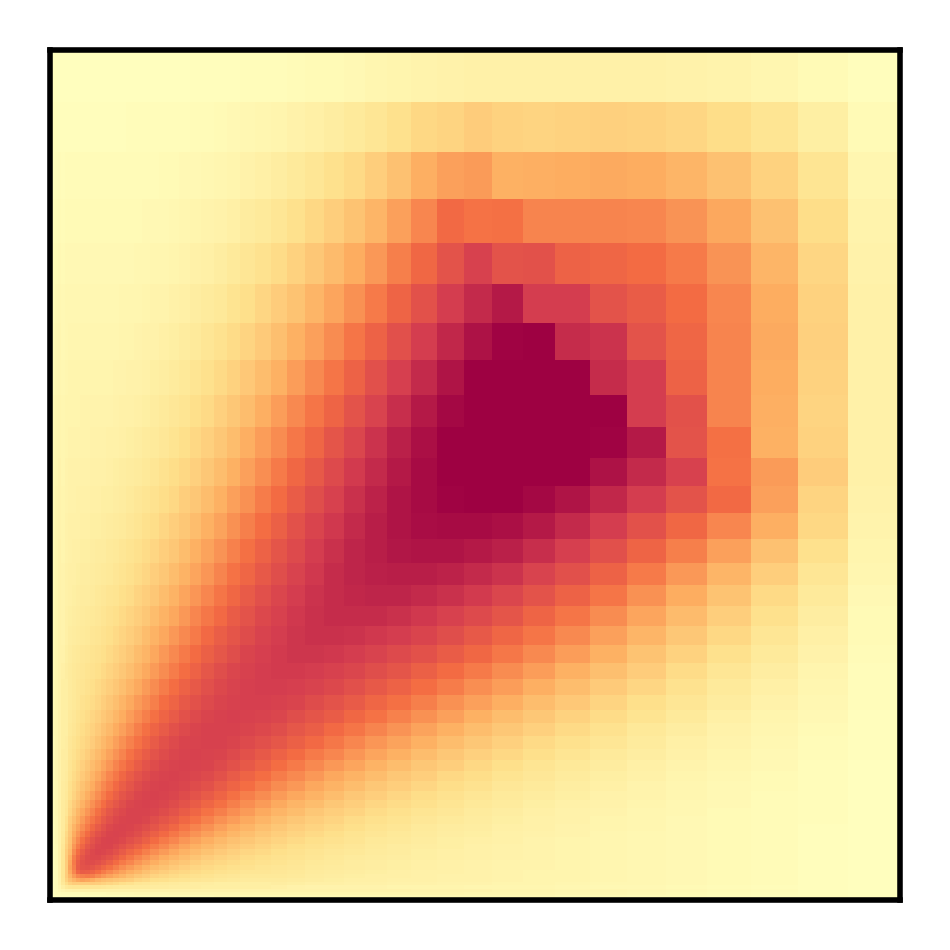}} & 
                \raisebox{-.5\height}{\includegraphics[scale=0.35, trim=10 0 10 0, clip]{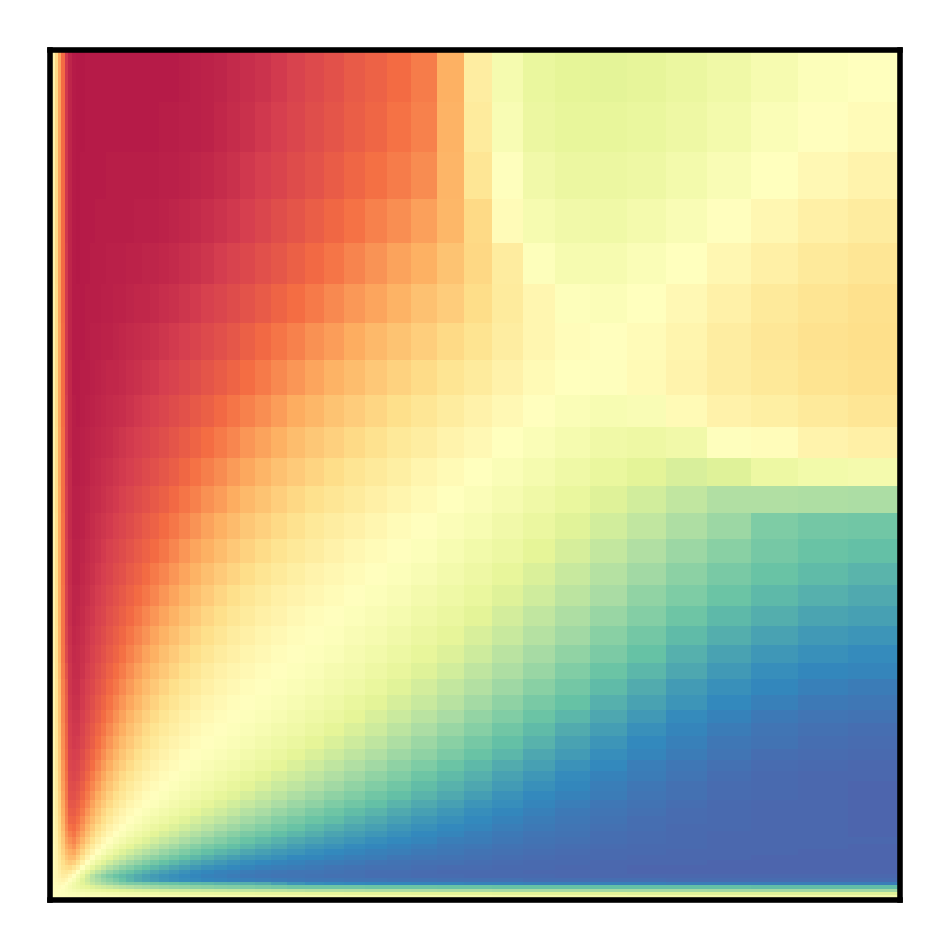}} 
                \\
                \rotatebox[origin=c]{90}{Truth} & 
                \raisebox{-.5\height}{\includegraphics[scale=0.35, trim=10 0 10 0, clip]{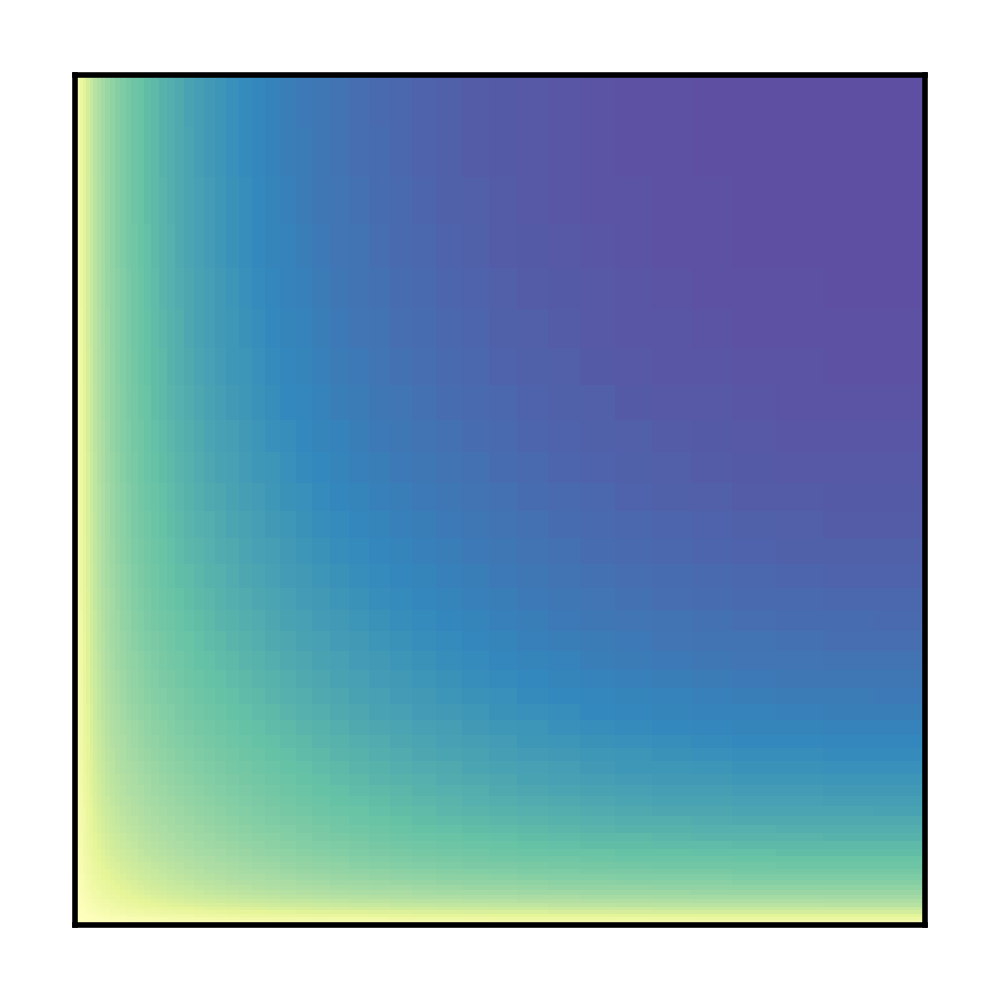}} & 
                \raisebox{-.5\height}{\includegraphics[scale=0.35, trim=10 0 10 0, clip]{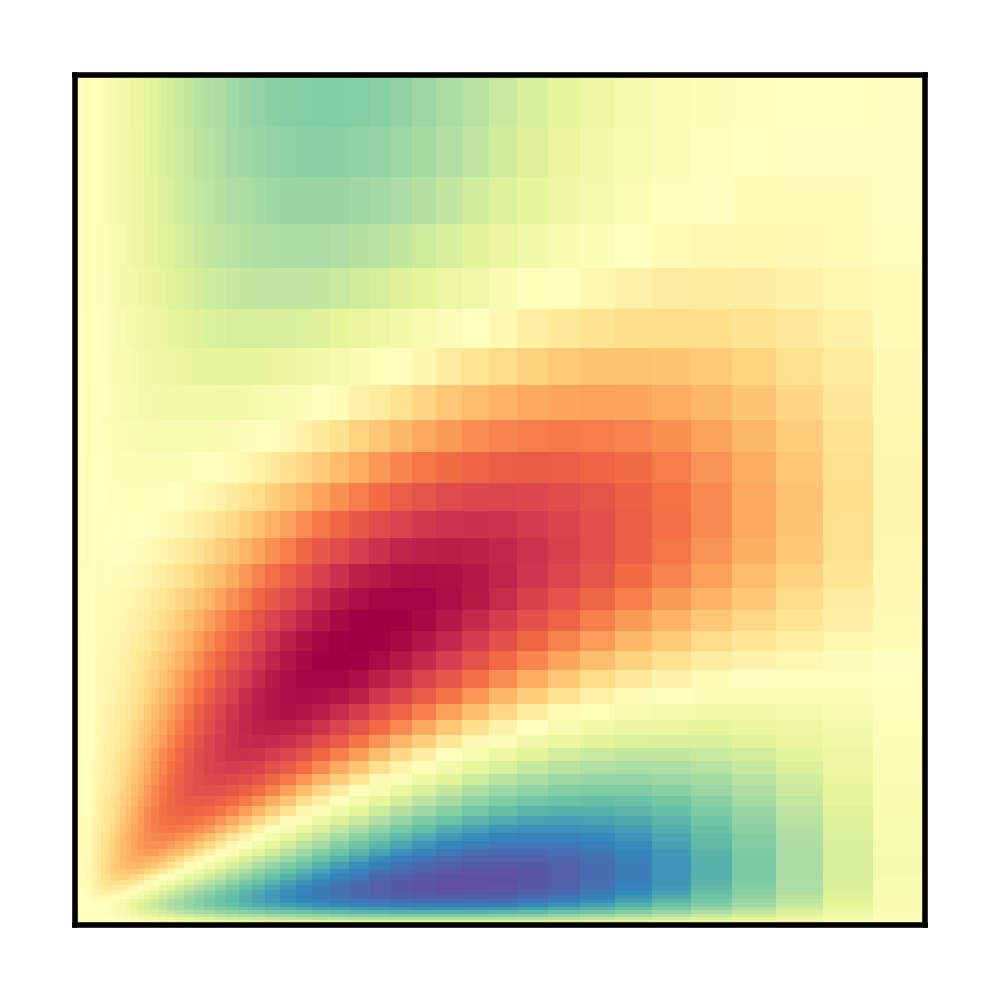}} & 
                \raisebox{-.5\height}{\includegraphics[scale=0.35, trim=10 0 10 0, clip]{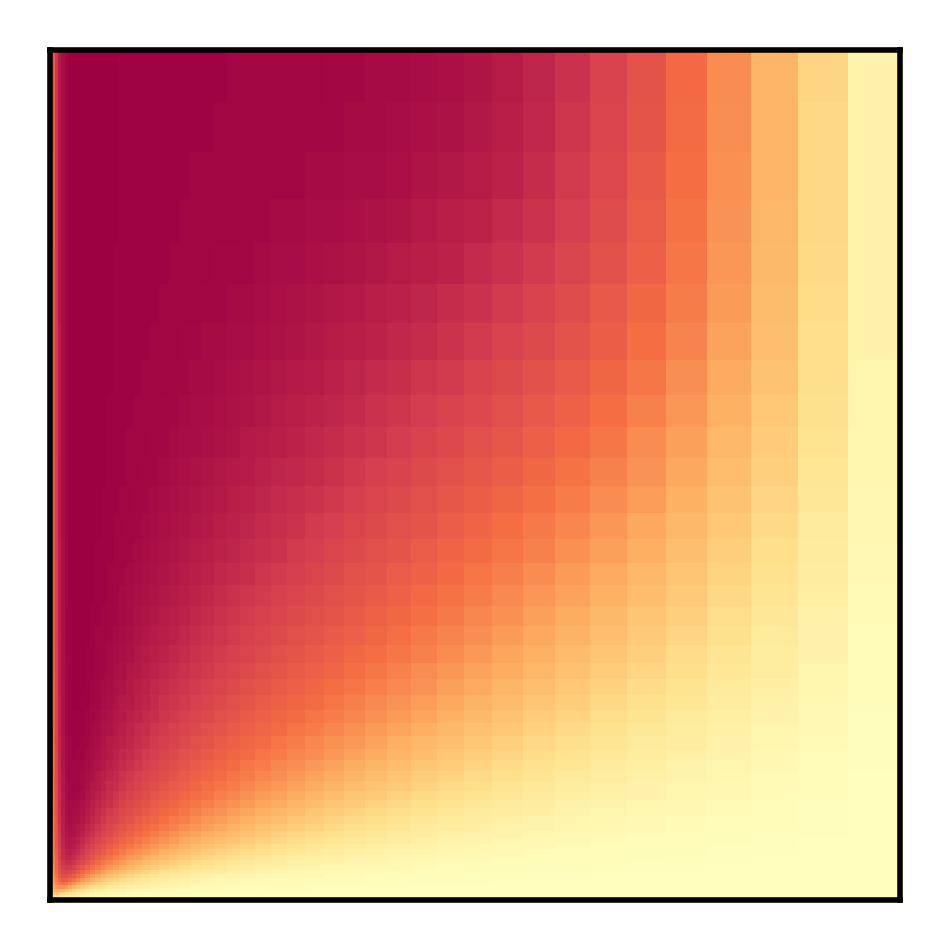}} & 
                \raisebox{-.5\height}{\includegraphics[scale=0.35, trim=10 0 10 0, clip]{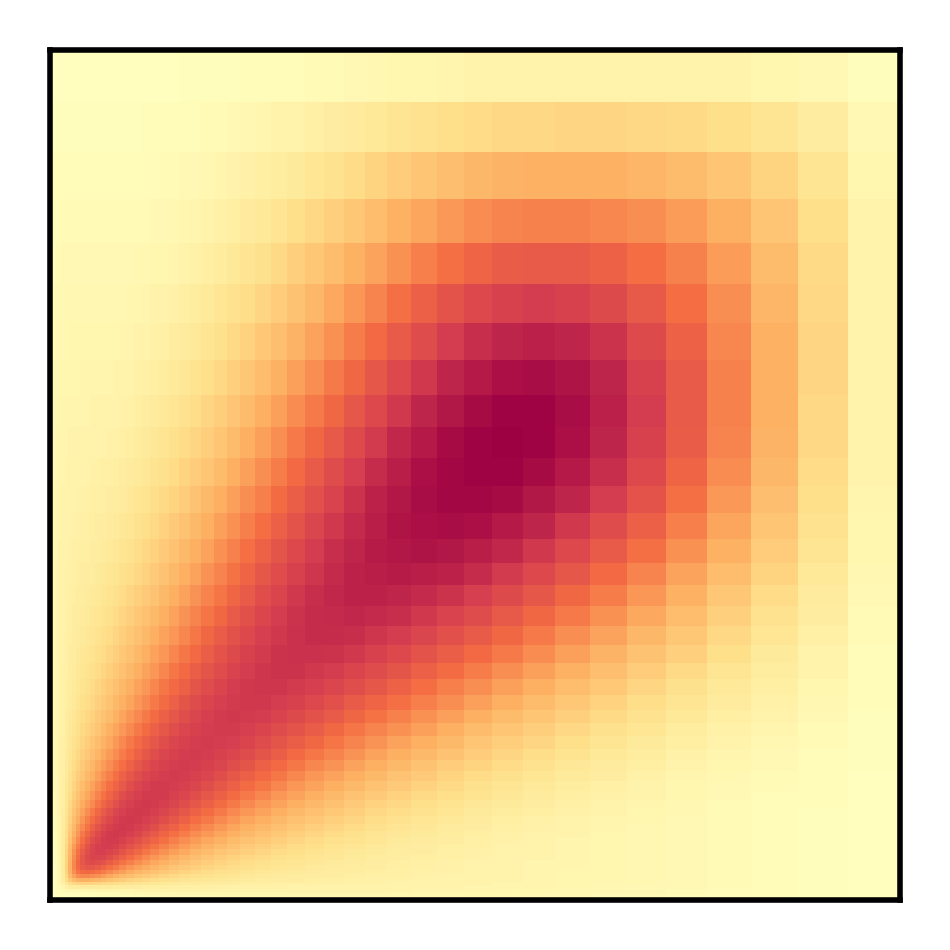}} & 
                \raisebox{-.5\height}{\includegraphics[scale=0.35, trim=10 0 10 0, clip]{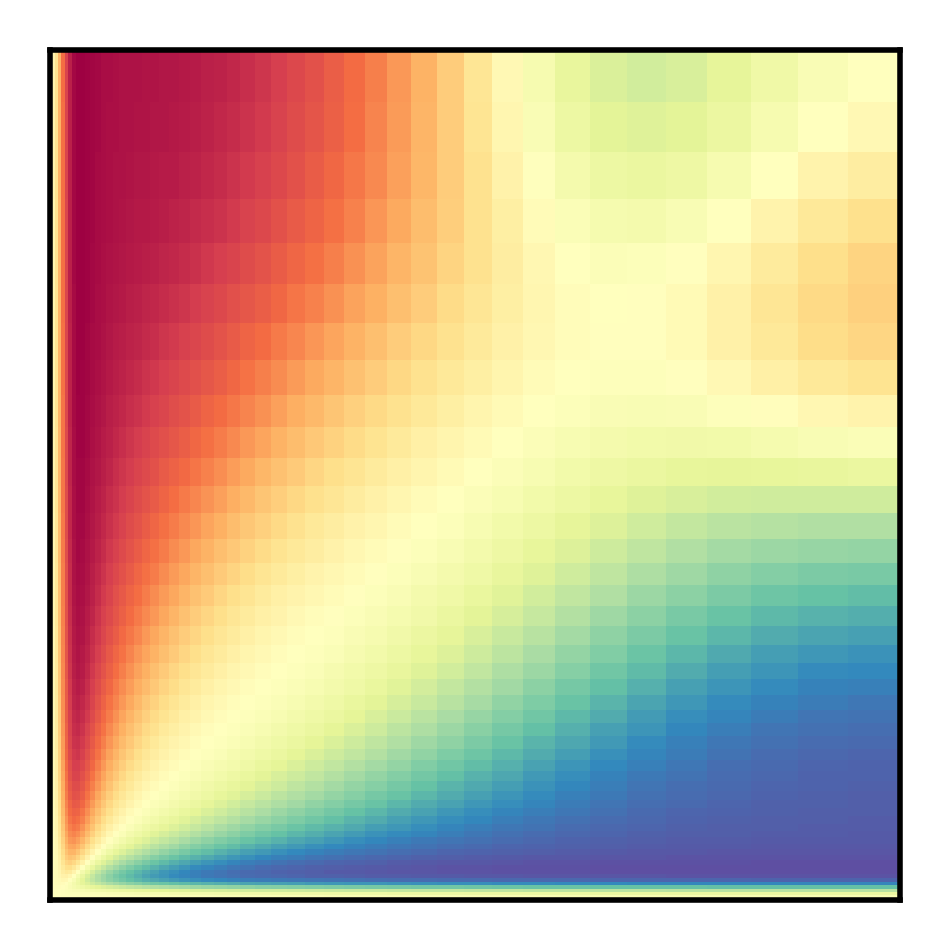}} 
                \\
                & \rotatebox[origin=l]{-90}{\includegraphics[trim=6 5 7 5, clip]{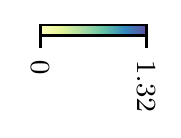}} &
                \rotatebox[origin=l]{-90}{\includegraphics[trim=6 5 7 5, clip]{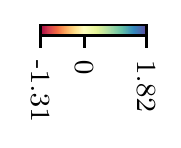}} &
                \rotatebox[origin=l]{-90}{\includegraphics[trim=6 5 7 5, clip]{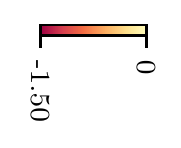}} &
                \rotatebox[origin=l]{-90}{\includegraphics[trim=6 5 7 5, clip]{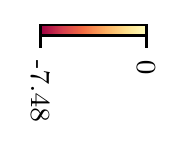}} &
                \rotatebox[origin=l]{-90}{\includegraphics[trim=6 5 7 5, clip]{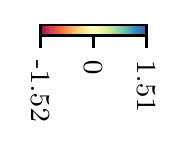}} 
            \end{tabular}
            }\hfill
            (\textit{b})\raisebox{-2.1\height}{ \raisebox{-.5\height}{\input{figure_4_b.tikz}}}
            \caption{\textit{(a)}: Velocity and anisotropic Reynolds stress results of learning a NLEVM from full field velocity measurements in flow through a square duct. The $u_z$ and $\reynoldstressbcomp_{xz}$ fields are the reflection of $u_y$ and $\reynoldstressbcomp_{xy}$ along the diagonal. \textit{(b)}: Schematic of flow through a square duct showing the secondary in-plane velocities. The simulation domain (bottom left quadrant) is highlighted. }
            \label{fig:duct_U_tau}
        \end{figure}

        The neural network has two inputs and four outputs and was pre-trained to the LEVM with $g^{(1)}=-0.09$.
        The turbulent equations used are those from the Shih quadratic $k$--$\varepsilon$ model. 
        Figure~\ref{fig:duct_g} shows the learned model which shows improved agreement with the truth. 
        The combination $g^{2}-0.5g^{(3)}+0.5g^{(4)}$ shows good agreement only for the higher range of scalar invariant $\theta_1$.
        This is due to the smaller scalar invariants corresponding to smaller velocity gradients and smaller magnitudes of the tensors $\Tt$.
        The velocity field is therefor expected to be less sensitive to the value of the Reynolds stress in these regions.
        It was observed that the smaller range of the invariant, where the learned model fails to capture the truth, occurs mostly in the center channel. 
        Figure~\ref{fig:duct_U_tau} shows the ability of the learned model to capture the correct velocity, including predicting the in-plane velocities, and the Reynolds stress. 
        The trained model fails to predict the correct $\reynoldstresscomp_{yz}$ in the center channel, but this does not propagate to the predicted velocities. 
        Additionally, it was observed that obtaining significant improvement in the velocity field requires only a few tens of training steps and only requires the coefficients to have roughly the correct order of magnitude. 
        On the other hand obtaining better agreement of the scalar coefficients took 1--2 orders of magnitude more training steps with diminishing returns in velocity improvement. 
        This shows the importance of using synthetic data to evaluate the ability of a training framework to learn the true underlying model when one exists rather than only comparing the quantities of interest.

    \subsection{Learning a LEVM from realistic data of flow over periodic hills}
        \label{sec:cases:extrapolation}
\begin{figure}
            \centering
            \raisebox{-.5\height}{\includegraphics[trim=11 0 10 0, clip]{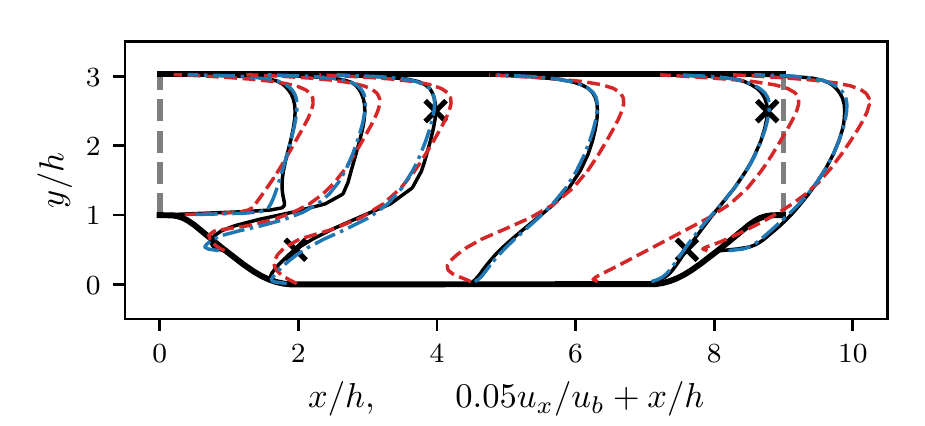}} \hfill
            \raisebox{-.5\height}{\includegraphics[trim=8 47 8 4, clip]{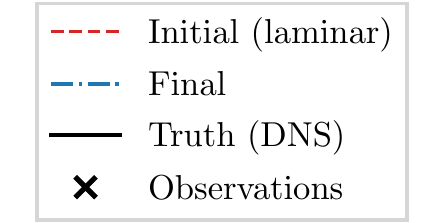}}
        \caption{Velocity results of learning an eddy viscosity model from sparse velocity data of flow over periodic hills. Six different profiles of $u_x$ velocity are shown. The training case corresponds to the baseline geometry $\alpha=1$.}
        \label{fig:hills_train}
        \end{figure}

        As a final test case, a LEVM is trained using sparse velocity measurements from DNS of flow over periodic hills. 
        The DNS data comes from \citet{xiao2020flows} who performed DNS of flow over periodic hills of varying slopes. 
        This flow is characterised by a recirculation region on the leeward side of the hill and scaling the hill width (scale factor $\alpha$) modifies the slope and the characteristics of the recirculation region (e.g. from mild separation for  $\alpha=1.5$ to massive separation for $\alpha=0.5$).
        For all flows, the Reynolds number based on hill height $h$ and bulk velocity $u_b$ through the vertical profile at the hill top is $Re=5,600$.  
        The training data consists of four point measurements of both velocity components in the flow over periodic hills with the baseline $\alpha=1$ geometry. 
        The two components of velocity are scaled equally in the objective function. 
        The training data and training results are shown in figure~\ref{fig:hills_train}. 
        The neural network in this case has one input and one output and is pre-trained to laminar flow, i.e. $g^{(1)}=0$. 
        The trained model is a spatially varying LEVM $g^{(1)}=g^{(1)}(\theta_1)$ that closely predicts the true velocity in most of the flow with the exception of the free shear layer in the leeward side of the hills. 

        \begin{figure}
            \centering
            \includegraphics[]{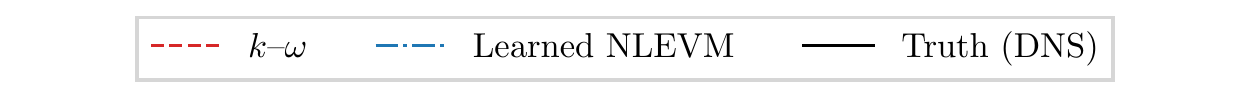} \\
            \vspace{-0.2em}
            (\textit{a})
            \raisebox{-1\height}
            {\includegraphics[trim=92 10 115 0, clip]{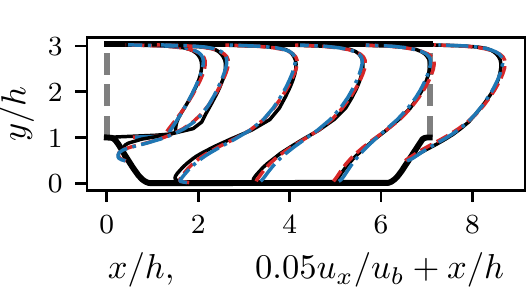}}
            (\textit{b})
            \raisebox{-1\height}
            {\includegraphics[trim=83 10 83 0, clip]{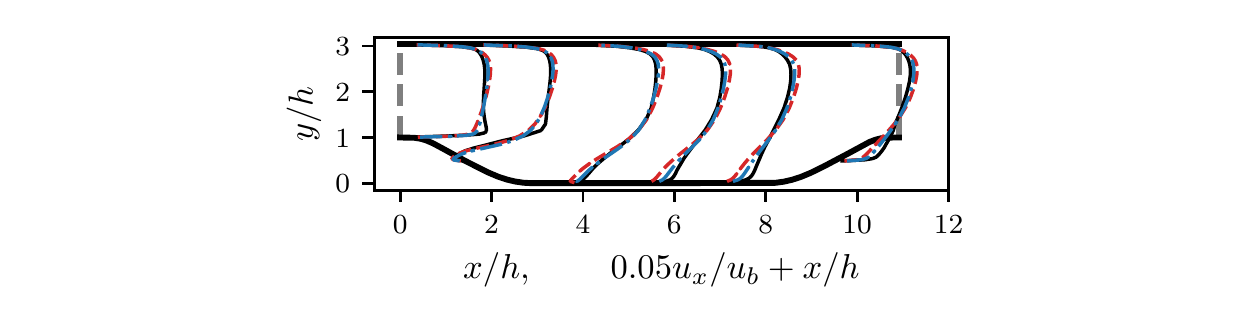}} 
        \caption{Comparison of horizontal velocity $u_x$ predictions using the trained eddy viscosity model $g^{(1)}=g^{(1)}(\theta_1)$ and the $k$--$\omega$ model $g^{(1)}=-0.09$ on two periodic hills geometries: (\textit{a}) $\alpha=0.5$ and (\textit{b}) $\alpha=1.5$. }
        \label{fig:hills_extrapolation}
        \end{figure}
        
        To test the extrapolation performance of the trained LEVM, we use it to predict the flow over the other periodic hill geometries, $\alpha\in[0.5, 0.8, 1.2, 1.5]$, and compare them to results with the $k$--$\omega$ model $g^{(1)}=-0.09$. 
        The results for $\alpha=0.5$ and $\alpha=1.5$ are shown in figure~\ref{fig:hills_extrapolation}. 
        For the $\alpha>1.0$ cases the trained linear model outperforms the $k$--$\omega$ model in the entire flow. 
        For the $\alpha<1.0$ cases the the trained model results in better velocity predictions in some regions, particulary the upper channel, while the $k$--$\omega$ model results in better velocities in the lower channel. 

\section{Conclusions}
    \label{sec:conclusions}
    In this paper we present a framework to train deep learning turbulence models using quantities derived from velocity and pressure that are readily available for a wide range of flows. 
    The method was first tested using synthetic data obtained from two traditional closure models: the linear $k$--$\omega$ and the Shih quadratic $k$--$\varepsilon$ models. 
    These two cases demonstrate the ability to learn the true underlying turbulence closure from measurement data when one exists. 
    The method was then used to learn a linear eddy viscosity model from synthetic sparse velocity data derived from DNS simulations of flow over periodic hills. 
    The trained model was used to predict flow over periodic hills of different geometries. 
    
    This work demonstrates that deep learning turbulence models can be trained from indirect observations when the relevant sensitivities of the RANS equations are available. 
    With the growing interest in differentiable programming for scientific simulations, it is expected that the availability of derivative information will become more commonplace in scientific and engineering computations, making it more seamless to couple scientific computations with novel deep learning methods.  
    
\section*{Acknowledgements}
    This material is based on research sponsored by the U.S. Air Force under agreement number FA865019-2-2204.
    The U.S. Government is authorised to reproduce and distribute reprints for Governmental purposes notwithstanding any copyright notation thereon.

\section*{Declaration of interests}
    The authors report no conflict of interest.

\bibliographystyle{jfm}
\bibliography{references}

\end{document}

%% file: figure_1.tikz
\begin{tikzpicture}
\newcommand{\fighalfwidth}{7}
\newcommand{\fighalfheight}{3.1}
\newcommand{\blockwidth}{6.15}
\newcommand{\blockheight}{3.0}
\newcommand{\blockhspacing}{5.0}
\newcommand{\txtys}{-3}
\newcommand{\ysleftdraw}{-2}
\newcommand{\ysrightdraw}{-2}
\tikzset{
    boxstyle/.style={rounded corners=10, Black!75, thick,},
}
\newcommand{\nnyshift}{3}
\newcommand{\nodesep}{4}
\newcommand{\layersep}{8}
\newcommand{\nodesize}{3}
\newcommand{\nhiddenlayers}{2}
\newcommand{\nhiddennodes}{4}
\newcommand{\nionodes}{2}
\tikzset{
    nn_layout/.style={node distance=\nodesep mm and \layersep mm},
    nn_node/.style={circle, minimum size=\nodesize mm, draw=black!75, thin},
    nn_hnode/.style={nn_node, fill=black!5},  
    nn_inode/.style={nn_node, fill=SpringGreen!50},
    nn_onode/.style={nn_node, fill=Apricot!50},
    nn_connection/.style={very thin, draw=black!75},
}
\newcommand{\arrowlength}{0.45*\blockwidth}
\newcommand{\arrowheight}{1.5}
\newcommand{\arrowhead}{1.5}
\newcommand{\arrowvspace}{2*\arrowhead}
\tikzset{
    arrowfwdstyle/.style={draw=Black!75, thick, single arrow, inner sep = \arrowheight mm, single arrow head extend= \arrowhead mm, text width = \arrowlength cm, align=center, shading=axis, left color=SpringGreen!20, right color=White}, 
    arrowbwdstyle/.style={draw=Black!75, thick, single arrow, inner sep = \arrowheight mm, single arrow head extend= \arrowhead mm, text width = \arrowlength cm, align=center, shading=axis, right color=Apricot!20, left color=White, shape border rotate=180}, 
}
\newcommand{\varhspace}{2}
\newcommand{\varvspace}{0}
\newcommand{\Layer}[4]{
    \pgfmathtruncatemacro{\ys}{(#3-2)*(\nodesep/2)} 
    \pgfmathtruncatemacro{\xs}{(#2+0.5)*\layersep} 
    \begin{scope}[yshift=\ys mm]
        \coordinate (#1_1) at (\xs mm, 0);
        \node [#4] (#1_c_1) at (#1_1) {};
        \foreach \y in {2,...,#3}
        {
            \pgfmathtruncatemacro{\yp}{\y-1}
            \coordinate [below=of #1_\yp] (#1_\y);
            \node [#4] (#1_c_\y) at (#1_\y) {};
        }
    \end{scope}
}
\newcommand{\FullyConnect}[4]{
    \foreach \x in {1,...,#2}{
        \foreach \y in {1,...,#4}
        {
            \draw [nn_connection] (#1_c_\x.east) -- (#3_c_\y.west);
        }
    }
}
\newcommand{\neuralnet}{
    \begin{scope}[nn_layout, xshift=-(\nhiddenlayers+2)*0.5*\layersep mm, yshift=\nodesep*0.5+\nnyshift mm]
        \Layer{input}{0}{\nionodes}{nn_inode}
        \foreach \xx in {1,...,\nhiddenlayers}{
            \Layer{hidden_\xx}{\xx}{\nhiddennodes}{nn_hnode}
        }
        \Layer{output}{\nhiddenlayers+1}{\nionodes}{nn_onode}
        \FullyConnect{input}{\nionodes}{hidden_1}{\nhiddennodes}
        \pgfmathtruncatemacro{\yh}{\nhiddenlayers-1}
        \foreach \yy in {1,...,\yh}{
            \pgfmathtruncatemacro{\ypo}{\yy+1}
            \FullyConnect{hidden_\yy}{\nhiddennodes}{hidden_\ypo}{\nhiddennodes}
        }
        \FullyConnect{hidden_\nhiddenlayers}{\nhiddennodes}{output}{\nionodes}
    \end{scope}
}
\pgfmathtruncatemacro{\lblockxshift}{\blockhspacing+\blockwidth*10} 
\pgfmathtruncatemacro{\blockyshift}{\fighalfheight*10-\blockheight*10}
\pgfmathtruncatemacro{\xs}{\blockwidth*10/2}
\pgfmathtruncatemacro{\ystxt}{\blockheight*10+\txtys}
\pgfmathtruncatemacro{\ysrdraw}{\blockheight*10/2+\ysrightdraw}
\pgfmathtruncatemacro{\ysldraw}{\blockheight*10/2+\ysleftdraw}
\pgfmathtruncatemacro{\llmx}{-\blockhspacing-\blockwidth*10/2}
\pgfmathtruncatemacro{\rlmx}{\blockhspacing+\blockwidth*10/2}
\coordinate (leftlowermid) at (\llmx mm, \blockyshift mm);
\coordinate (rightlowermid) at (\rlmx mm, \blockyshift mm);
\begin{scope}[xshift=-\lblockxshift mm, yshift=\blockyshift mm]
    \draw[boxstyle] (0, 0) rectangle (\blockwidth cm, \blockheight cm) {};
    \begin{scope}[xshift=\xs mm, yshift=\ysldraw mm]
        \neuralnet
        \begin{scope}[xshift=-19 mm,  yshift=3 mm]
            \node {$\boldsymbol{\theta}$};
        \end{scope}
        \begin{scope}[xshift=19 mm,  yshift=3 mm]
            \node {$\boldsymbol{g}$};
        \end{scope}
        \begin{scope}[xshift=0 mm,  yshift=10.0 mm]
            \node {$\boldsymbol{w}$};
        \end{scope}
    \end{scope}
    \begin{scope}[xshift=\xs mm, yshift=4 mm]
        \node [text width=\blockwidth cm, align=center] {$\boldsymbol{\mathcal{T}}(k,t_\tau)=\boldsymbol{0}$};
    \end{scope}
    \begin{scope}[xshift=\xs mm, yshift=\ystxt mm]
        \node [text width=\blockwidth cm, align=center] {Turbulence Model};
    \end{scope}
    \node [arrowfwdstyle, below=\arrowvspace mm of leftlowermid] (tlarrow) {Neural Network};
    \node [arrowbwdstyle, below=\arrowvspace mm of tlarrow] (blarrow) {Backpropagation};
\end{scope}
\pgfmathtruncatemacro{\ysarrow}{\blockyshift+\blockheight*10/2+\ysleftdraw}
\pgfmathtruncatemacro{\xspacearrow}{\blockhspacing-0.2}
\draw [-{latex}, Black, thick] (-\xspacearrow mm, \ysarrow mm) -- node [above, midway] {$\reynoldstress$} (\xspacearrow mm, \ysarrow mm);
\begin{scope}[xshift=\blockhspacing mm, yshift=\blockyshift mm]
    \draw[boxstyle] (0, 0) rectangle (\blockwidth cm, \blockheight cm) {};
    \begin{scope}[xshift=\xs mm, yshift=\ysrdraw mm]
        \node [text width=\blockwidth cm, align=center] (rans)
        { 
        $\begin{array}{c} 
        \boldsymbol{u\cdot\nabla u} - \mathrm{\nu}\boldsymbol{\nabla^2u} + \boldsymbol{\nabla\cdot} \textcolor{black}{\reynoldstress} + \nabla p - \boldsymbol{s} = \boldsymbol{0} \\[4pt]
        \boldsymbol{\nabla\cdot u} = 0 \\[8pt]
        J = J(\boldsymbol{u}, p)
        \end{array}$
        };
    \end{scope}
    \begin{scope}[xshift=\xs mm, yshift=\ystxt mm]
        \node [text width=\blockwidth cm, align=center] {Obervation Operator};
    \end{scope}
    \node [arrowfwdstyle, below=\arrowvspace mm of rightlowermid] (trarrow) {PDEs};
    \node [arrowbwdstyle, below=\arrowvspace mm of trarrow] (brarrow) {Adjoint Model};
\end{scope}
\node [left = \varhspace mm of blarrow, align=center] (leftvarnode) {\Large$\frac{\partial J}{\partial \boldsymbol{w}}$};
\node [right = \varhspace mm of trarrow] (rightvarnode) {$J$};
\coordinate (leftvar) at (leftvarnode);
\coordinate (rightvar) at (rightvarnode);
\node [align=center] at ($(leftvar |- tlarrow.west)$) {$\boldsymbol{w}$};
\node at ($(tlarrow.east -| 0,0)$) {$\reynoldstress$};
\node [right = \varhspace mm of trarrow] {$J$};
\node at ($(blarrow.east -| 0,0)$) {\Large$\frac{\partial J}{\partial \reynoldstress}$};
\end{tikzpicture} 

%% file: figure_4_b.tikz
\begin{tikzpicture}
\newcommand{\squarelength}{20} 
\newcommand{\harrowshift}{-2} 
\newcommand{\coordshift}{3} 
\newcommand{\coordlength}{10} 
\definecolor{tab-blue}{HTML}{4e79a7}
\tikzset{
    outline/.style={draw=black!50, opacity=1.0,},
    midprof/.style={draw=black},
    symm/.style={dashed, black!50},
    measurement/.style={{latex}-{latex},},
    coord/.style={-{latex}, very thin},
    inplaneflow/.style={black!50, thin},
}
\newcommand{\inplane}[3]{
        \coordinate (Corner) at (#1); 
        \coordinate (Mid) at ($(#1) !0.75! (#2) $); 
        \coordinate (Center) at ($(#1) !0.5! (#3) $); 
        \tikzmath{%
          coordinate \P, \I;
          \P1 = (Corner); \P2 = (Mid); \P3 = (Center);
          \a = veclen(\Px3-\Px2, \Py3-\Py2);
          \b = veclen(\Px1-\Px3, \Py1-\Py3);
          \c = veclen(\Px2-\Px1, \Py2-\Py1);
          \m = max(\a, \b, \c);
          \a = \a / \m; \b = \b / \m; \c = \c / \m;
          \ix = (\a*\Px1 + \b*\Px2 + \c*\Px3) / (\a + \b + \c);
          \iy = (\a*\Py1 + \b*\Py2 + \c*\Py3) / (\a + \b + \c);
          \I = (\ix, \iy);
        }
        \pgfmathsetmacro{\ix}{\ix*1pt/1cm} 
        \pgfmathsetmacro{\iy}{\iy*1pt/1cm} 
        \coordinate (O) at (\ix cm, \iy cm);
        \path (Corner); \pgfgetlastxy{\Ax}{\Ay};
        \path (Mid); \pgfgetlastxy{\Bx}{\By};
        \path (Center); \pgfgetlastxy{\Cx}{\Cy};
        \foreach[evaluate={\t=1-~}] ~ in {0.1, 0.3, 0.5}
        \draw[inplaneflow, scale around={~:(O)}, smooth cycle, tension=\t]
            plot coordinates {($(\Ax, \Ay)$) ($(\Bx, \By)$) ($(\Cx, \Cy)$)};
    }
\path (0,0) coordinate (ai) -- (0,\squarelength mm) coordinate [pos=0.5] (wi) coordinate (bi) -- (\squarelength mm, \squarelength mm) coordinate [pos=0.5] (ni) coordinate (ci) -- (\squarelength mm, 0) coordinate [pos=0.5] (ei) coordinate (di) -- (0,0) coordinate [pos=0.5] (si); 
\path (ai) -- (ci) coordinate [pos=0.5] (mi);
\fill[tab-blue!25] (mi) -- (si) -- (ai) -- (wi);
\inplane{ai}{mi}{wi}
\inplane{ai}{mi}{si}
\inplane{bi}{mi}{wi}
\inplane{bi}{mi}{ni}
\inplane{ci}{mi}{ni}
\inplane{ci}{mi}{ei}
\inplane{di}{mi}{ei}
\inplane{di}{mi}{si}
\draw [symm] (ni) -- (si);
\draw [symm] (wi) -- (ei);
\draw [symm] (ai) -- (ci);
\draw [symm] (bi) -- (di);
\draw [coord] (ai) -- ($(di)+(0:\coordshift mm)$) coordinate (xe); 
\node (x1) [anchor=west] at (xe) {$z$};
\draw [coord] (ai) -- ($(bi)+(90:\coordshift mm)$) coordinate (xn); 
\node (x1) [anchor=south] at (xn) {$y$};
\draw[midprof] (ai) -- (bi) -- (ci) -- (di) -- cycle; 
\draw [measurement] ($(ai)+(90:\harrowshift mm)$) -- ($(di)+(90:\harrowshift mm)$) node [midway, below] {$2h$};
\end{tikzpicture}